\documentclass{article}

\usepackage{amssymb,amsfonts,amsmath,stmaryrd}
\usepackage{cite,enumerate,float,indentfirst}
\usepackage{color}

\def\be{\begin{eqnarray}}
\def\ee{\end{eqnarray}}

\def\Tr{{\rm Tr}\,}

\def\Gns{{\rm Gns}}

\newcommand{\beq}{\begin{equation}}
\newcommand{\eeq}{\end{equation}}
\newcommand{\beqa}{\begin{eqnarray}}
\newcommand{\eeqa}{\end{eqnarray}}

\definecolor{red}{rgb}{1,0,0}
\definecolor{orange}{rgb}{1,0.5,0}
\definecolor{violet}{rgb}{0.7,0,1}

\hoffset= - 1.0in         

\begin{document}
\title{\bf
Elliptic $q,t$ matrix models
}

\author{
{\bf Andrei Mironov$^{a,b,c}$}\footnote{mironov@lpi.ru; mironov@itep.ru},
\ and  \  {\bf Alexei Morozov$^{d,b,c}$}\thanks{morozov@itep.ru}
}
\date{ }

\maketitle

\vspace{-5.0cm}

\begin{center}
\hfill FIAN/TD-09/20\\
\hfill IITP/TH-16/20\\
\hfill ITEP/TH-25/20\\
\hfill MIPT/TH-14/20
\end{center}

\vspace{2.cm}

\begin{center}
$^a$ {\small {\it Lebedev Physics Institute, Moscow 119991, Russia}}\\
$^b$ {\small {\it ITEP, Moscow 117218, Russia}}\\
$^c$ {\small {\it Institute for Information Transmission Problems, Moscow 127994, Russia}}\\
$^d$ {\small {\it MIPT, Dolgoprudny, 141701, Russia}}
\end{center}

\vspace{.0cm}

\begin{abstract}
The Gaussian matrix model is known to deform to the $q,t$-matrix model. We consider further deformation to the elliptic $q,t$ matrix model by properly deforming the Gaussian density as well as the Vandermonde factor. Properties of an associated basis of symmetric functions that provide the matrix model property $<char>\sim {\rm char}$ in the deformed elliptic case are discussed.
\end{abstract}

\section{Introduction}

Eigenvalue
matrix models  are associated with many different issues in modern string theory.
Basically, they appear in two different forms:
\begin{itemize}
\item models in Miwa variables with logarithimic potential, like
\be
\int\prod_i dz_i\prod_{i\ne j}(z_i-z_j)\prod_{i,a}(z_i-x_a)^{\mu_a}
\ee
\item and the standard matrix models of the Gaussian type in time variables \cite{UFN3}, like
\be
\int\prod_i dz_i\prod_{i\ne j}(z_i-z_j)\prod_{i}e^{-z_i^2/2+\sum_kp_kz_i^k/k}
\ee
\end{itemize}
Integrals (1) and (2) {\it per se}
are directly related by Miwa transform $p_k = \sum_a x_a^k$,
but things become more involved after various deformations of logarithms,
exponentials and integrals.
The logarithmic models attract a lot of attention these days,
because they are associated with both the AGT correspondence \cite{AGT}
and with CFT/Painlev\'e correspondence \cite{Rain},
where one applies peculiar techniques based on
various Selberg and Kadell type formulas \cite{Sel},
while, in the case of ordinary time-variables, one
has direct relation to KP/Toda integrability and Virasoro-like constraints,
which get obscure after reduction to a few Miwa variables.

On the technical side, matrix models of these both types are associated with systems of symmetric polynomials. In particular, matrix models in Miwa variables are associated with the Jack polynomials \cite{Sel}, or their generalized versions \cite{gJ}. There is also a $q,t$-deformation of these matrix model, which is associated with the Macdonald \cite{MMSS} and generalized Macdonald \cite{gM} polynomials. Moreover, a further elliptic extension of {\it some} of Selberg type formulas was also recently proposed \cite{Spi}.

As for the matrix models in time variables, a reference example of such a model is the Gaussian Hermitian matrix model. A connection of such a matrix model with a set of symmetric functions goes the following way:
 a defining feature of these matrix models is their superintegrability property \cite{MM,MMsr}, which claims that the average of a properly chosen symmetric function is proportional to ratios of symmetric functions on a proper loci: $<char>\sim {\rm char}$. In the particular case of the ordinary $N\times N$ Gaussian Hermitian model, this symmetric function is an $SL(N)$ character, the Schur polynomial.
 
 An immediate deformation of the Gaussian Hermitian model to the $q,t$ matrix model was recently presented in \cite{MPSh}. This does not come as a surprise that the set of symmetric polynomials associated with such deformation turns out to be Macdonald polynomials. Technically, such an approach proved to be a very powerful tool to construct deformations of systems of symmetric polynomials. Hence, the next important step to be done is to construct an elliptic deformation of the $q,t$ matrix model and an associated system of symmetric polynomials.

Indeed, as soon as the Schur polynomials associated with the Gaussian Hermitian model are eigenfunctions of the Calogero Hamiltonains, while the Macdonald polynomials associated with the $q,t$ matrix model are the eigenfunctions of the trigonometric Ruijsenaars Hamiltonians, one may expect that the symmetric functions associated with the elliptic $q,t$ model are the eigenfunctions of the double elliptic Hamiltonians (or, rather, of the Hamiltonians dual to the elliptic Ruijsenaars Hamiltonains).  Recently, we proposed \cite{AKMMdell2,AKMMdell3} candidates for this role, and the subject is very topical. One of the problems is that, despite there is a candidate for the role of the eigenfunctions, there are no Hamiltonians for them! Indeed, the naive Hamiltonians proposed in \cite{KSh,GZ} have distinct eigenfunctions (see \cite[sec.6.3]{AKMMdell1}). It is still an open question if these two systems of functions can be related by a unitary transform.

In this paper, we will try to make the first step: to introduce a proper definition of the elliptic deformation of $q,t$ matrix model. The problem of constructing a corresponding set of symmetric polynomials will be reported elsewhere (see first results in \cite{MMellloci}).

\paragraph{Notation.}

We use the notation for the Schur polynomials as polynomials of symmetric variables $z_i$: $S_R(z_i)$ and as polynomials of power sums $p_k:=\sum_iz^k_i$: $S_R\{p_k\}$ and similarly for other symmetric polynomials.

We define the $q$-Pocchammer symbol:
\be
(z;q)_\infty:=\prod_{i=0}^\infty(1-q^iz);\ \ \ \ \ \ \ (z;q,w)_\infty:=\prod_{i,j=0}^\infty(1-q^iw^jz);\ \ \ \ \ \ \ (z;q)_n:={(z;q)_\infty\over (zq^n;q)_\infty};\ \ \ \ \ \ \
\ee
We also define
$$
\delta^{(n)}_k:=\sum_{j=0}^{n-1}e^{2\pi i{kj\over n}}
$$
The Jackson integralo is defined
\be
\int_{-a}^af(z)d_qz:=a(1-q)\sum_{i=0}^\infty q^i\Big(f(aq^i)+f(-aq^i)\Big)
\ee

\section{Standard matrix models}

\subsection{Hermitean Gaussian matrix model}

The standard Hermitean Gaussian matrix model is defined by the partition function
\be\label{H1}
Z_N(p_k):={1\over V_N}\int dH\exp\Big(-{1\over 2}\Tr H^2+\sum_k {p_k\over k}\Tr H^k\Big)
\ee
where $H$ is the Hermitean $N\times N$ matrix, $dH$ is the Haar measure on the Hermitean matrices, $V_N$ is volume of the unitary group $U(N)$, and the integral is understood as a formal power series in variables $t_k$.

This partition function is a generating function of all gauge-invariant correlators
\be\label{H2}
\left< \prod_i\Tr H^{k_i}\right>:={1\over Z_N(0)}\int dH \prod_i\Tr H^{k_i}\exp\Big(-{1\over 2}\Tr H^2\Big)
\ee

One can also integrate over the angular variables in (\ref{H1}) in order to obtain \cite{Mehta}
\be\label{He1}
Z_N(p_k)={1\over N!}\int_{-\infty}^\infty \prod_i dz_i \prod_{j\ne i}(z_i-z_j)\exp\Big(-{1\over 2}\sum_iz_i^2+\sum_{i,k} {p_k\over k}z_i^k\Big)
\ee
and
\be
\left< \prod_i \sum_mz_m^{k_i}\right>:={1\over Z_N(0)}\int_{-\infty}^\infty \prod_i dz_i \prod_{j\ne i}(z_i-z_j)
\left(\prod_i \sum_mz_m^{k_i}\right) \exp\Big(-{1\over 2}\sum_iz_i^2\Big)
\ee
where $z_i$ are the eigenvalues of $H$.
To deal with the later integral, one can use the moments of the density $\rho^H(z)$:
\be
{\int_{-\infty}^\infty dz\rho(z)z^{2k}\over \int_{-\infty}^\infty dz\rho(z)}=(2k-1)!!,\ \ \ \ \ \ \ \ \rho(z):=\exp\Big(-{1\over 2}z^2\Big)
\ee
The gauge invariant correlators (\ref{H2}) are correlators of symmetric functions in terms of matrix eigenvalues $z_i$. Hence, in order to completely solve the model, one needs to know all correlators of any complete set of symmetric functions. Here we notice that the basis of Schur functions $S_R(z_i)$ is the most convenient one, and that
\be\label{S}
\Big< S_R(z_i)\Big>={S_R\{\delta_{k,2}\}\cdot S_R\{N\}\over S_R\{\delta_{k,1}\}}
\ee
This is the formula that provides the complete solution of the model.

\subsection{$q,t$-matrix model}

The $q,t$-matrix model is defined \cite{MPSh} directly in terms of the eigenvalue integral (\ref{He1}),
\be\label{qt1}
Z_N^{(q,t)}(p_k)=\int_{-\xi}^\xi
\left(\prod_i z_i^{\beta(N-1)}\rho^{(q)}(z_i)d_qz_i\right) \prod_{j\ne i}\left({z_i\over z_j};q\right)_\beta
\exp\Big(\sum_{i,k} {p_k\over k}z_i^k\Big)
\ee
and
\be\label{qt2}
\left< \prod_i \sum_mz_m^{k_i}\right>_{(q,t)}:={1\over Z_N^{(q,t)}(0)}\int_{-\xi}^\xi \left(\prod_i z_i^{\beta(N-1)}\rho^{(q)}(z_i)d_qz_i\right) \prod_{j\ne i}\left({z_i\over z_j};q\right)_\beta
\left(\prod_i \sum_mz_m^{k_i}\right)
\ee
where $t:=q^\beta$, $\xi$ is a parameter, the $q$-Pocchammer symbol is defined as
\be
(z;q)_\infty:=\prod_{k=0}^{\infty}(1-q^kz),\ \ \ \ \ \ \ (z;q)_\beta:={(z;q)_\infty\over (zq^\beta;q)_\infty}
\ee
the integral is defined to be the Jackson integral,
\be
\int_{-\xi}^\xi d_qz f(z):=\xi(1-q)\sum_{k=0}^\infty q^k\Big(f(\xi q^k)+f(-\xi q^k)\Big)
\ee
and the $q$-deformed density
\be
\rho^{(q)}(z):=(q^2z^2/\xi^2;q^2)_\infty
\ee
gives the Gaussian density $\rho(z):=\exp\Big(-{1\over 2}z^2\Big)$ in the limit of $q\to 1$.

\subsection{Performing Jackson integral}

Now let us see that this $q$-deformation of the density leads to the most natural $q$-deformation of the density moments.

We use the Andrews-Askey formula \cite[Eq.(2.10.20)]{GR}
\be\label{AA}
\int_{-\xi}^\xi{(q^2z^2/\xi^2;q^2)_\infty\over (-\mu_1z/\xi;q)_\infty(\mu_2z/\xi;q)_\infty}d_qz=\xi (1-q)
{(q^2;q^2)_\infty (-1;q)_\infty (\mu_1\mu_2;q)_\infty\over (\mu_1^2;q^2)_\infty (\mu_2^2;q^2)_\infty}
\ee
in the particular case of $\mu_1=\mu_2=\mu$:
\be
\int_{-\xi}^\xi{(q^2z^2/\xi^2;q^2)_\infty\over (\mu^2z^2/\xi^2;q^2)_\infty}d_qz=\xi(1-q)\prod_{n=0}^\infty {(1-q^{2n+2})(1+q^n)
(1-q^n\mu^2)\over (1-q^{2n}\mu^2)^2}
\ee
Since in the particular case of $\mu=0$, it gives
\be\label{norm}
\int_{-\xi}^\xi\rho^{(q)}(z)d_qz=\xi(1-q)\prod_{n=0}^\infty (1-q^{2n+2})(1+q^n)
\ee
one obtains
\be
\left<{1\over (\mu^2z^2/\xi^2;q^2)_\infty}\right>=\prod_{n=0}^\infty {1-q^n\mu^2\over (1-q^{2n}\mu^2)^2}
\ee
where
\be
\Big<f(z)\Big>:={\int_{-\xi}^\xi\rho^{(q)}(z)f(z)d_qz\over \int_{-\xi}^\xi\rho^{(q)}(z)d_qz}
\ee
It also allows one to get moments of $\rho(z)$ by expansion in $\mu^2$: using
\be
{1\over (\mu^2z^2/\xi^2;q^2)_\infty}=\sum_{m=0}^\infty{1\over (q^2;q^2)_m}\left({\mu z\over\xi}\right)^{2m}
\ee
and
\be\label{Pe}
(z;q)_\infty=\exp\left[-\sum_m{z^m\over (1-q^m)m}\right]
\ee
one obtains
\be\label{aux}
\left<{1\over (\mu^2z^2/\xi^2;q^2)_\infty}\right>=\sum_{m=0}^\infty{1\over (q^2;q^2)_m}\left({\mu\over\xi}\right)^{2m}\Big<z^{2m}\Big>=
\exp\left[\sum_m{\mu^{2m}\over m}\left({2\over 1-q^{2m}}-{1\over 1-q^m}\right)\right]
\ee
This gives
\be\label{av}
\Big<z^k\Big>={1\over 2}\xi^{k}\cdot\delta^{(2)}_k\cdot\prod_{i=1}^{k/2}\Big(1-q^{2i-1}\Big)
\ee
As soon as $\xi$ enters trivially as a scaling factor, from now on, we put $\xi=1$.

\subsection{Matrix model averages}

Thus, we could see that the $q$-deformation of the density just replaces the double factorials by the quantum double factorials. Now we can calculate the correlators in the $q,t$-model (\ref{qt1}). The most natural basis of symmetric functions in this case is given by the Macdonald polynomials $M_R(z_i)$, and one gets similarly to (\ref{S}):
\be\label{M}
\Big<M_R(z_i)\Big>={M_R\Big({[Nk]\over[k]}\Big)\cdot
M_R\Big({\delta^{(2)}_k\over 1-t^k}\Big)\over M_R\Big({1\over 1-t^k}\Big)}
\ee
In particular, at $N=1$, the only non-zero Macdonald polynomial is $M_{[r]}(z)=z^r$, and there is an identity (see  (\ref{av}))
\be\label{measure}
\Big<M_{[r]}(z)\Big>=\Big<z^{r}\Big>={M_{[r]}\Big({\delta^{(2)}_k\over 1-t^k}\Big)\over M_{[r]}\Big({1\over 1-t^k}\Big)}
\ee
which is quite non-trivial, because the both Macdonald polynomials at the r.h.s. of this formula do depend on $t$, but the ratio does not!

In order to prove (\ref{measure}), one can multiply its both parts with ${(t;q)_k\over (q;q)_k}\lambda^rM_{[r]}\Big({1\over 1-t^k}\Big)$
and use the Cauchy formula,
\be
\sum_k {(t;q)_k\over (q;q)_k}\lambda^kM_{[k]}\{p_k\}=
\sum_k {(t;q)_k\over (q;q)_k}M_{[k]}(\lambda)M_{[k]}\{p_k\}=
\exp\left(\sum_{m=1}{1-t^m\over 1-q^m}{\lambda^mp_m\over m}\right)
\ee
and (\ref{Pe}) in order to see that (\ref{measure}) is equivalent to the identity
\be
\left<{1\over (z\lambda ;q)_\infty}\right>={1\over (\lambda^2;q^2)_\infty}
\ee
This identity is just the Andrews-Askey integral (\ref{AA}) at $\mu_1=0$, $\mu_2=\lambda$.

\section{Elliptic $q,t$ matrix models}

\subsection{The definition}

Let us define the $\theta$-function\footnote{It differs by a factor from the standard odd $\theta$-function:
\be
\theta_1(u;\tau)={iw^{1/8}\cdot (w;w)_\infty\over \sqrt{z}}\cdot\theta_{w}(z)\Big|_{w=e^{2\pi i\tau},z=e^{2\pi i u}}
\ee}
\be
\theta_w(z):=(z;w)_\infty (w/z;w)_\infty
\ee
and the elliptic $\Gamma$-function
\be
\Gamma(z;w,q):={(qw/z;w,q)_\infty\over (z;w,q)_\infty}=\exp\left[\sum_m{z^m-(wq/z)^m\over (1-q^m)(1-w^m)m}\right]
\ee
In particular,
\be
\Gamma(q^n;w,q)=\prod_{k=1}^\infty{1-w^k\over 1-q^k}\cdot\prod_{i=1}^{n-1}\theta_w(q^i)
\ee

Now, in order to define an elliptic $q,t$ matrix models, one has to define two points: (i) a proper elliptic generalization of the measure (\ref{measure}), and (ii) a proper elliptic generalization of the Vandermonde factor.
In order to generalize (\ref{measure}), one can naturally {\it define} it from the generalization of (\ref{av}) as
\be\label{ave}
\boxed{
\Big<z^k\Big>_{(ell)}\sim\xi^{k}\cdot\delta^{(2)}_k\cdot\prod_{i=1}^{k/2}\theta_w(q^{2i-1})
}
\ee
while the Vandermonde factor is given just by the elliptic $\Gamma$-function. Thus, finally, we define (compare with \cite{NZ} at $t=q$, i.e. $\beta=1$) 
\be\label{ell1}
Z_N^{ell}(p_k)=\int
\left(\prod_i z_i^{\beta(N-1)}\rho^{(ell)}(z_i)d_{ell}z_i\right) \prod_{j\ne i}{\Gamma\left(q^\beta{z_i\over z_j};w,q\right)\over
\Gamma\left({z_i\over z_j};w,q\right)}
\exp\Big(\sum_{i,k} {p_k\over k}z_i^k\Big)
\ee
and
\be\label{ell2}
\boxed{
\left< \prod_i \sum_mz_m^{k_i}\right>_{(ell)}:={1\over Z_N^{(ell)}(0)}\int \left(\prod_i z_i^{\beta(N-1)}\rho^{(ell)}(z_i)d_{ell}z_i\right) \prod_{j\ne i}{\Gamma\left(q^\beta{z_i\over z_j};w,q\right)\over
\Gamma\left({z_i\over z_j};w,q\right)}
\left(\prod_i \sum_mz_m^{k_i}\right)
}
\ee
Here the yet-to-be-defined integral $\int\ldots d_{ell}z$ is technically fixed by formula (\ref{ave}). 

\subsection{Factorization property}

There is now an important step to be done: one has to find a counterpart $E_R$ of the Macdonald polynomials that would provide the proper basis with property (\ref{M}), corresponding deformations of the special loci also have to be realized:
\be\label{E}
\Big<E_R(z_i)\Big>_{(ell)}={E_R\Big(p^*_k(N)\Big)\cdot
E_R\Big(\Big[1+(-1)^k\Big]p_k^{\Delta^{\rm ell}}\Big)\over E_R\Big(p_k^{\Delta^{\rm ell}}\Big)}
\ee
In particular, at $N=1$, there should be an identity
\be\label{Emeasure}
\Big<E_{[r]}(z)\Big>_{(ell)}=\Big<z^{r}\Big>_{(ell)}={E_{[r]}\Big(\Big[1+(-1)^k\Big]p_k^{\Delta^{\rm ell}}\Big)\over E_{[r]}\Big(p_k^{\Delta^{\rm ell}}\Big)}
\ee
One could expect that these polynomials are the elliptic GNS \cite{AKMMdell2,AKMMdell3} (generalization of  the Noumi-Shiraishi polynomials from \cite{NS}), this still has to be checked. Note that there are two sets of the elliptic GNS polynomials orthogonal to each other, and which one of them would emerge is an open question. Since of these two sets, the set of conjugate polynomials create a ring consistent with the tensor product of representations of $SL(N)$ (i.e. the corresponding generalized Littlewood-Richardson coefficients do not vanish iff they do not vanish for the corresponding Schur functions), they are most plausible candidate.

The first few examples of the elliptic GNS polynomials are:
\be
\Gns_{[1]} = p_1, \ \ \ \ \ \ 
\Gns_{[2]} = \left(1-\frac{\zeta(q,t,w)}{2}\right)p_2 + \zeta(q,t,w)\frac{p_1^2}{2},
\ \ \ \ \ \ 
\Gns_{[1,1]} = \frac{-p_2+p_1^2}{2}
\ee
while those of the conjugate ones are:
\be
\Gns_{[1]}^\perp = p_1, \ \ \ \ \ \ 
\Gns_{[2]}^\perp = \frac{p_2+p_1^2}{2}, \ \ \ \ \ \ 
\Gns_{[1,1]}^\perp = p_1^2-\zeta(q,t,w)\cdot\frac{p_2+p_1^2}{2}
\ee
where
\be
\zeta(q,t,w):={\theta_w(q^2)\theta_w(t)\over\theta_w(qt)\theta_w(q)}
\ee

Note that (\ref{Emeasure}) does not depend on $t$ (see (\ref{ave})), since all the dependence on $t$ enters through the Vandermonde factor. Hence, one may expect that the factorization properties related with (\ref{measure}), would also persist at $t=q$, when the polynomials $E_R$ reduce to the Schur polynomials. In other words, one has to find an elliptic deformation of the loci that preserves the factorization property, in particular, in the Schur polynomials. It can be actually done, and will be reported elsewhere \cite{MMellloci}.

\section{Conclusion}   

To conclude, we suggested two different definitions of elliptic matrix model: explicit eigenvalue integral (\ref{ell2}) and explicit prescription for time-dependent characters  averages (\ref{E}). Usually it takes quite some time to establish an exact relation/equivalence between such two formulation, and we do not try to do it in the present short note. Still, we hope that it will not take too long, since an elliptic model is badly needed in modern studies of AGT relations and DIM algebras.  For the first steps in this direction, see \cite{MMellloci}.

\section*{Acknowledgements}

This work was supported by the Russian Science Foundation (Grant No.20-12-00195).


\begin{thebibliography}{12}

\bibitem{AGT} R. Dijkgraaf, C. Vafa, arXiv:0909.2453;\\
H. Itoyama, K. Maruyoshi, T. Oota,
Prog.Theor.Phys. {\bf 123} (2010) 957-987, arXiv:0911.4244\\
T. Eguchi, K. Maruyoshi,
arXiv:0911.4797;
arXiv:1006.0828\\
R. Schiappa, N. Wyllard,
arXiv:0911.5337\\
A. Mironov, A. Morozov, Sh. Shakirov,
JHEP {\bf 02} (2010) 030, arXiv:0911.5721;
Int.J.Mod.Phys. {\bf A25} (2010) 3173-3207, arXiv:1001.0563\\
H. Itoyama, T. Oota, Nucl. Phys. {\bf B838} (2010) 298-330, arXiv:1003.2929\\
A. Mironov, A. Morozov, An. Morozov, Nucl.Phys. {\bf B843} (2011) 534-557, arXiv:1003.5752

\bibitem{Rain} A.Mironov, A.Morozov, 	
Phys.Lett. {\bf B773} (2017) 34-46,  arXiv:1707.02443; Phys.Lett. {\bf B785} (2018) 207-210, arXiv:1708.07479\\
A.~Mironov, A.~Morozov and Z.~Zakirova,
  JHEP {\bf 1910} (2019) 227,
  arXiv:1908.01278

\bibitem{Sel} A.~Mironov, A.~Morozov and S.~Shakirov,
  JHEP {\bf 1102} (2011) 067,
  arXiv:1012.3137

\bibitem{UFN3} A. Morozov,
Phys.Usp.(UFN) {\bf 37} (1994) 1;
hep-th/9502091; hep-th/0502010\\
A. Mironov, Int.J.Mod.Phys. {\bf A9} (1994) 4355; Phys.Part.Nucl.
{\bf 33} (2002) 537; hep-th/9409190

\bibitem{gJ} V. A. Alba, V. A. Fateev, A. V. Litvinov and G. M. Tarnopolskiy, Lett. Math. Phys. {\bf 98} (2011) 33
arXiv:1012.1312\\
A. Belavin and V. Belavin, Nucl. Phys. {\bf B850} (2011) 199, arXiv:1102.0343\\
A. Morozov, A. Smirnov, Lett.Math.Phys. {\bf 104} (2014) 585-612, arXiv:1307.2576\\
S.~Mironov, A.~Morozov and Y.~Zenkevich,
  JETP Lett.\  {\bf 99} (2014) 109,
  arXiv:1312.5732

\bibitem{MMSS} A.~Mironov, A.~Morozov, S.~Shakirov and A.~Smirnov,
  Nucl.\ Phys.\ B {\bf 855} (2012) 128,
  arXiv:1105.0948

\bibitem{gM} Y.~Zenkevich,
  JHEP {\bf 1505} (2015) 131,
  arXiv:1412.8592

\bibitem{Spi} V.~P.~Spiridonov,
  "Partition Functions and Automorphic Forms'', Lecture Notes of the
  Dubna Winter School (Eds. V.A. Gritsenko, V.P. Spiridonov), Moscow Lectures,
  vol. 5. Springer, Heidelberg (2020) 271-318,
  arXiv:1912.12971

\bibitem{MM} A.~Mironov and A.~Morozov,
  Phys.\ Lett. {\bf B771} (2017) 503,
  arXiv:1705.00976

\bibitem{MMsr} A.~Mironov and A.~Morozov,
  JHEP {\bf 1808} (2018) 163,
  arXiv:1807.02409

\bibitem{MPSh} A.~Morozov, A.~Popolitov and S.~Shakirov,
  Phys.\ Lett. {\bf B784} (2018) 342,
  arXiv:1803.11401

\bibitem{AKMMdell2} H.~Awata, H.~Kanno, A.~Mironov and A.~Morozov,
  JHEP {\bf 2008} (2020) 150,
  arXiv:2005.10563
  
\bibitem{AKMMdell3} H.~Awata, H.~Kanno, A.~Mironov and A.~Morozov,
  Eur.\ Phys.\ J. {\bf C80} (2020)  994,
  arXiv:2002.12746
 
\bibitem{KSh}   P.~Koroteev, S.~Shakirov,
Lett.Math.Phys. (2019),
arXiv:1906.10354

\bibitem{GZ} A. Grekov, A. Zotov, arXiv:2010.08077

\bibitem{AKMMdell1}   H.~Awata, H.~Kanno, A.~Mironov, A.~Morozov,
JHEP {\bf 2020} (2020) 212,   arXiv:1912.12897

\bibitem{MMellloci} A. Mironov, A. Morozov, arXiv:2011.02855

\bibitem{GR}  G. Gasper, M. Rahman, {\it Basic hypergeometric series},
Cambridge University Press, 1990

\bibitem{NZ} A.~Nedelin and M.~Zabzine,
  JHEP {\bf 1703} (2017) 098,
  arXiv:1511.03471

\bibitem{NS} M. Noumi, J. Shiraishi, 
arXiv:1206.5364

\end{thebibliography}
\end{document}